\documentclass[11pt]{article}

\usepackage{amscd}
\usepackage{amsmath}

\begin{document}

\newcommand{\N}{N\raise.7ex\hbox{\underline{$\circ $}}$\;$}

\thispagestyle{empty}

\begin{center}
{\bf

\thispagestyle{empty}

BELARUS NATIONAL ACADEMY OF SCIENCES

B.I. STEPANOV's  INSTITUTE OF PHYSICS

}
\end{center}

\vspace{40mm}

\begin{center}

{\bf  V.S. Otchik,  V.M.  Red'kov \footnote{E-mail:
redkov@dragon.bas-net.by}, }

\end{center}

\vspace{5mm}

\begin{center}

{\bf ELECTROMAGNETIC WAVES IN THE DE SITTER SPACE
 }

\end{center}

\thispagestyle{empty}

 \begin{quotation}

5-Dimensional  wave equation for a massive particle of spin 1 in
the background of de Sitter  space-time model is solved in static
coordinates. The spherical 5-dimensional vectors $A_{a}, a=
1,...,5$ of three types, $j,j+1, j-1$ are constructed. In massless
case  they give electromagnetic wave solutions, obeying the
Lorentz condition. 5-form of equations in massless  case  is used
to produce recipe to build  electromagnetic wave solutions of the
types $\Pi, E,M$; the first is trivial and can be removed by a
gauge transformation. The recipe  is specified to produce
spherical $\Pi, E, M$ solutions in static coordinates.

\end{quotation}

\vspace{10mm}


 electric and magnetic waves, gauge symmetry

{\bf 33C05;  34B05}

\vspace{20mm}
 This paper is based on the old one:

\vspace{10mm}

 V.S. Otchik, V.M.  Red'kov.

    Spherical waves of electric, magnetic and longitudinal types in de Sitter space.

    Minsk  (1986).  44 pages. Deposited in VINITI  16.12.86,  8641 - B86 (in Russian)
\footnote{All-Russian Scientific and Technical Information
Institute of Russian Academy of Sciences -
  Vserossiisky Institut Nauchnoi i Tekhnicheskoi Informatsii (VINITI).}

\newpage

\section{Introduction}

Examining  fundamental  particle fields on the background of
expanding universe, in particular de Sitter and anti de Sitter
models, has a long history    [1-30]. Special value of these
geometries consists in their simplicity and high symmetry groups
underlying them which makes us to believe in existence of exact
analytical treatment for  some fundamental problems of classical
and quantum field theory  in curved spaces. In particular, there
exist special representations for fundamental wave equations,
Dirac's and Maxwell's, which are explicitly invariant under
respective symmetry groups $SO(4.1)$  and $SO(3.2)$ for these
models. The interest to exact solutions of wave equations for
particles with different  spins in de Sitter  space remains steady
and inexhaustible [1-30].

  In present paper the wave equation in 5-dimensional  form for a massive particle of spin 1 in the background of
de Sitter  space-time model is solved in static coordinates $(t,
r, \theta, \phi )$, covering the part of space-time to  event
horizon. The spherical 5-dimensional waves $A_{a}(t, r, \theta,
\phi ), a= 1,...,5$ of three types, $j,j+1, j-1$ are constructed.
In massless case  they gives electromagnetic wave solutions,
obeying the Lorentz condition. Group-theoretical 5-dimensional
form of equations in massless  case  is used to produce recipe to
build  electromagnetic wave solutions of the types $\Pi, E,M$; the
first is trivial and can be removed by a  gauge transformation.
The recipe is applicable in arbitrary coordinates, in the paper it
is specified in   static coordinates of the de Sitter space.

\section{On the  5-theory of a massive spin 1 particle in de Sitter space
}

It is known that a wave equation for spin 1 field on the
background of de Sitter space-time can be presented in a form
explicitly invariant under the group  $SO(4.1)$. To specify some
details of that approach, let start with covariant Proca equations
\begin{eqnarray}
\nabla _{\alpha} \Psi _{\beta } \; - \; \nabla _{\beta } \Psi
_{\alpha } = m \; \Psi _{\alpha \beta } \; , \qquad \nabla ^{\beta
} \Psi _{\alpha \beta } = m \; \Psi _{\alpha }\; , \nonumber
\label{13.1.1}
\end{eqnarray}

\noindent from whence  it follows equation for the vector
$\Psi_{\alpha }$
\begin{eqnarray}
( \nabla ^{\beta } \; \nabla _{\beta } +  m^{2} )\Psi _{\alpha } -
\nabla _{\alpha } ( \nabla ^{\beta }  \Psi _{\beta } ) -
R_{\alpha \beta }  \Psi ^{\beta }  = 0\; . \label{13.1.2}
\end{eqnarray}

\noindent Because  $\Psi _{\alpha }$  obeys the Lorentz condition
$
 \nabla ^{\beta }  \Psi _{\beta } = 0$, eq. (\ref{13.1.2}) gives
\begin{eqnarray}
(\; \nabla ^{\beta } \nabla _{\beta }  +\; m^{2} ) \;\;
\Psi^{\beta}
  -\; R_{\alpha \beta }\;\Psi ^{\beta }  = 0 \; .
\nonumber \label{13.1.4}
\end{eqnarray}

\noindent In the  massless case, instead of  (\ref{13.1.1}) we
have
\begin{eqnarray}
\nabla _{\alpha }\; \Psi _{\beta } \;-\; \nabla _{\beta } \Psi
_{\alpha } = \Psi _{\alpha \beta } \; , \qquad \nabla ^{\beta }
\Psi _{\alpha \beta }  = 0\; ; \nonumber \label{13.1.5}
\end{eqnarray}

\noindent and and the second order equation is
\begin{eqnarray}
\nabla ^{\beta } \nabla _{\beta } \Psi^{\alpha}
  - \nabla _{\alpha } ( \nabla ^{\beta }
\Psi _{\beta }) - R_{\alpha \beta } \Psi ^{\beta }  = 0 \; .
\label{13.1.6}
\end{eqnarray}

\noindent Eq.  (\ref{13.1.5})  has a class of trivial (gauge)
solutions
\begin{eqnarray}
\tilde{\Psi }_{\alpha }  = \nabla _{\alpha } \; f  =
\partial _{\alpha } \; f \; \;, \qquad
\tilde{\Psi }_{\alpha \beta } = 0 \; , \label{13.1.7}
\end{eqnarray}

\noindent where   $f(x)$  is an arbitrary scalar function.
Commonly, this fact is linked up to the gauge principle
\begin{eqnarray}
\Psi _{\alpha }(x) \; \sim\;  \Psi _{\alpha }(x)  \;+\; \partial
_{\alpha }\; f(x)  \; . \nonumber \label{13.1.8}
\end{eqnarray}

\noindent Trivial solution  $\tilde{\Psi }_{\alpha }(x)$  obeys
the Lorentz condition if $f(x)$  satisfies
\begin{eqnarray}
 \nabla ^{\alpha } \nabla _{\alpha } \;  f(x) \equiv
  \Delta \; f(x) = 0 \; .
\nonumber \label{13.1.9}
\end{eqnarray}

Now let us specify the above equations in conformal  coordinates
in de Sitter space \cite{1973-Hawking-Ellis}
\begin{eqnarray}
dS^{2} = {1 \over \Phi^{2} } \; [ \; (dx^{0})^{2}  - (dx^{1})^{2}
- (dx^{2})^{2} - (dx^{3})^{2} \; ]\; , \nonumber
\\
\Phi  = (1 - x^{2})/2 \; , \;\; x^{2}  = (x^{0})^{2} - (x^{1})^{2}
- (x^{2})^{2} - (x^{3})^{2} \; ; \label{13.1.10}
\end{eqnarray}

\noindent   $x^{0} = c t/\rho, \ldots$ Bellow it will be
convenient to use coordinates
\begin{eqnarray}
x_{\alpha } = \eta _{\alpha \beta } x^{\beta } \; , \; \eta
_{\alpha \beta } = \mbox{diag} ( +1, -1, -1, -1 ) \; , \;
\nonumber
\\
g_{\alpha \beta} = {1 \over \Phi ^{2} } \eta _{\alpha \beta } \; ,
\;
\partial _{\alpha } \Phi  = - x_{\alpha } \;\; , \;\;
 \partial ^{\alpha } \equiv  \eta ^{\alpha \beta }
 \partial _{\beta }  \; .
\nonumber \label{13.1.11}
\end{eqnarray}

\noindent  Christoffel symbols are
\begin{eqnarray}
\Gamma ^{\rho }_{\alpha \beta } = {1 \over \Phi ^{2} } \; ( \;
\delta ^{\rho }_{\alpha } \; x_{\beta }  \; - \; \delta
^{\rho}_{\beta } \; x_{\alpha }  \; - \; x^{\rho } \eta _{\alpha
\beta }\; ) \; ; \nonumber
\end{eqnarray}

\noindent and Proca equations  take the form
\begin{eqnarray}
\partial _{\alpha } \Psi _{\beta }  - \partial _{\beta } \Psi _{\alpha } =
m\; \Psi _{\alpha \beta } \;, \qquad  \Phi ^{2} \; \partial
^{\beta } \Psi _{\alpha \beta } = m \Psi _{\alpha } \; .
\label{13.1.12}
\end{eqnarray}

\noindent In massless case we have
\begin{eqnarray}
\partial _{\alpha } \Psi _{\beta } - \partial _{\beta } \Psi _{\alpha } =
\Psi _{\alpha \beta } \; ,\qquad  \partial ^{\beta } \Psi _{\alpha
\beta } = 0 \; . \nonumber \label{13.1.13}
\end{eqnarray}

\noindent The Lorentz condition in these coordinates looks
\begin{eqnarray}
\partial ^{\alpha } \; \Psi _{\alpha }  = - \;{2 \over \Phi ^{2} } \;
x^{\alpha } \; \Psi _{\alpha } \; . \nonumber \label{13.1.14}
\end{eqnarray}

Now, starting with  $x^{\alpha }$,  let us introduce five
coordinates $\xi ^{a}$
\begin{eqnarray}
\xi ^{\alpha } = { x^{\alpha } \over \Phi  }\; , \qquad (\alpha  =
0  , 1, 2, 3) \; ,\qquad   \xi ^{5} = { 1 + x^{2} \over 1 - x^{2}}
\; , \nonumber
\\
x^{\alpha } = { \xi ^{\alpha } \over 1  + \xi ^{5} }\; , \qquad
\Phi  = { 1 \over 1  + \xi ^{5}} \; , \qquad  a = \alpha  ,\;   5
\;  ; \label{13.1.15a}
\end{eqnarray}

\noindent they are characterized by
\begin{eqnarray}
{ \partial  \xi ^{\alpha } \over \partial  x^{\beta }} = {1 \over
\Phi ^{2}} \; (\; \Phi  \; \delta ^{\alpha }_{\beta }\;+\;
x^{\alpha } x_{\beta }\; ) \; , \qquad  { \partial  \xi ^{5} \over
\partial  x^{\beta }} = {x_{\beta } \over \Phi ^{2}} \; ,
\nonumber
\\
 {
\partial  x^{\alpha } \over \partial  \xi ^{\beta }} = \Phi\;
\delta ^{\alpha }_{\beta } \; , \qquad  { \partial  x^{\alpha }
\over \partial  \xi ^{5}} = - \Phi \; x^{\alpha }\; ; \nonumber
\\
 (\xi ^{0})^{2} - ( \xi ^{1})^{2} - ( \xi ^{2})^{2} - ( \xi
^{3})^{2} - ( \xi ^{5})^{2}  = - 1 \;  . \nonumber
\\
dS^{2} = {1 \over \Phi ^{2} } \; \eta _{a b} \; dx^{a} dx^{b} =
\eta _{\alpha \beta } \; d\xi ^{\alpha } d\xi ^{\beta } \;-\;
(d\xi ^{5})^{2}\; . \label{13.1.16}
\end{eqnarray}

Therefore, de Sitter space can be identified with a sphere in
 5-dimensional pseudo-Euclidean space,
and thereby it has 10-parametric symmetry group $SO(4.1)$
\begin{eqnarray}
\xi ^{a'} = S^{a}_{\;\;b}\; \xi ^{b} \; , \qquad  ( S^{a}_{\;\;b})
\in SO(4.1) \; . \nonumber
 \end{eqnarray}

\noindent Instead of 4-vector  $\Psi ^{\alpha }(x)$ (below it will
be designated as
 $a^{\alpha }(x) )$ let us introduce 5-vector  $A^{a}(\xi)$
\begin{eqnarray}
A^{a}(\xi )  = { \partial \xi ^{a} \over \partial x^{\alpha } } \;
a^{\alpha }(x) \; , \nonumber
\\
 A^{\alpha }  = {1 \over \Phi ^{2}}
(\Phi   \delta ^{\alpha }_{\beta } + x^{\alpha }  x_{\beta }  )
a^{\beta }   \; , \qquad  A^{5} = {1 \over \Phi ^{2}} x_{\alpha }
a^{\alpha }  \;  ; \label{13.1.18}
\end{eqnarray}

\noindent The vector $A^{a}(\xi )$  transforms as a 5-vector $\xi
^{a}$  under the group $SO(4.1)$
\begin{eqnarray}
A ^{a'}(\xi ' )  = { \partial \xi  ^{a'} \over \partial x^{\alpha
} } \;a^{\alpha }(x) = [ \; {\partial  \over \partial x^{\alpha }
} ( S^{a}_{\;\;b} \; \xi ^{b})\; ]\; a^{\alpha }(x) =
S^{a}_{\;\;b} \; A^{b}(\xi ) \; . \label{13.1.19}
\end{eqnarray}

\noindent Inverse relationship  to  (\ref{13.1.18}) has the form
\begin{eqnarray}
a^{\alpha }(x)  = { \partial x^{\alpha } \over \partial\; \xi
^{a}}\; A^{a}  =  \Phi \; ( A^{\alpha }\;  - \; x^{\alpha }
A^{5})\;  . \nonumber \label{13.1.20}
\end{eqnarray}

\noindent Fife variables  $A^{a}(\xi )$  are not independent --
the following condition holds
\begin{eqnarray}
\xi ^{a}  A_{a}  = 0 \;  . \label{13.1.21}
\end{eqnarray}

\noindent  A wave equation  for the 5-vector  $A^{a}(\xi )$
invariant   under the group $SO(4.1)$
 should be constructed with the help of the following operator
\begin{eqnarray}
L_{ab}   =  \xi _{a}\; {\partial \over \partial \xi ^{b} }  - \xi
_{b} \; {\partial \over \partial \xi ^{a} }  \; , \nonumber
\label{13.1.22}
\end{eqnarray}

\noindent and its possible form is
\begin{eqnarray}
 {1\over 2} L^{ab} L_{ab} A_{c} \; + \; \kappa \;
L_{ca}\; A^{a}  \; +\;  \sigma \; A_{c}  = 0 \; , \label{13.1.23}
\end{eqnarray}

\noindent where  $\kappa $ and  $\sigma $ are constants. It is
readily verified that the Lorentz condition has the following
5-form
\begin{eqnarray}
L_{ab} A^{b}  = A_{a} \; ; \label{13.1.24}
\end{eqnarray}

\noindent therefore  eq.  (\ref{13.1.23})  looks
\begin{eqnarray}
 (\; {1 \over 2} L^{ab} L_{ab} \; + \; (\kappa \; + \; \sigma )
 \;  ) \; A_{c} = 0 \; .
\label{13.1.25}
\end{eqnarray}

\noindent Bearing in mind  (\ref{13.1.16}), one  finds
\begin{eqnarray}
L_{\alpha \beta }  =   x_{\alpha } {\partial \over \partial
x^{\beta }} \;-\; x_{\beta } {\partial  \over \partial x^{\alpha }
}\;  , \;\;\; L_{5\alpha } =  - \Phi  {\partial \over \partial
x^{\alpha }}\; +
 \; x^{\beta } \;  L_{\alpha \beta } \; ;
 \nonumber
 \label{13.1.26}
 \end{eqnarray}

\noindent and further
\begin{eqnarray}
{1 \over 2} \; L^{ab}\;  L_{ab}  = - \Phi ^{2} \; (\; \partial
^{\alpha } \partial _{\alpha }\;+\; 2 \; \Phi \; x^{\alpha }\;
\partial _{\alpha } \; ) \; .
\nonumber \label{13.1.27}
\end{eqnarray}

\noindent The later coincides with covariant d'Alamber  operator
in conformal coordinates
\begin{eqnarray}
{1 \over 2} \; L^{ab} \; L_{ab} = - \Delta \; . \nonumber
\label{13.1.28}
\end{eqnarray}

\noindent Comparing  (\ref{13.1.25}) with  (\ref{13.1.12}), we
find expression for $(\kappa  + \sigma )$
\begin{eqnarray}
(\; {1 \over 2}\; L^{ab} \; L_{ab}\; + \; m^{2}\; + \; 2\; ) \;
A_{c} = 0 \; . \label{13.1.29}
\end{eqnarray}

\noindent Setting here  $m^{2} = 0$,   we get the wave equation
for a massless field; also we should remember on eqs.
(\ref{13.1.21})  and   (\ref{13.1.24}). Let us derive 5-form for
the above trivial solution
\begin{eqnarray}
\tilde{a}_{\alpha } = {\partial  \over \partial x^{\alpha } }
f\;\; , \qquad  \Delta\; f = 0 \; , \nonumber
\end{eqnarray}

\noindent transforming it to 5-form
\begin{eqnarray}\tilde{A}_{\alpha }  =  (\Phi
{\partial  \over \partial x^{\alpha } }  +  x_{\alpha } x^{\beta
}\; {\partial  \over \partial x^{\beta } }) f \; ,\qquad
\tilde{A}_{5} =  - x^{\alpha } {\partial  \over
\partial x^{\alpha } }  f \; ,
\nonumber \label{13.1.30b}
\end{eqnarray}

\noindent or shortly
\begin{eqnarray}
\tilde{A}_{a} = (  {\partial  \over \partial \xi ^{a}} + \xi _{a}
\xi ^{b} {\partial  \over \partial \xi ^{b}} ) \; f \equiv  m_{a}
f \; . \label{13.1.30c}
\end{eqnarray}

\noindent In the following we will use an identity $ \Delta   =
m^{a} \; m_{a} \; , \; \;  m_{a}\; A^{a} = 0.$

\section{  Spherical waves in static coordinates, massive case}

Equations for a vector particle will be solved in static
coordinates in de Sitter space \cite{1973-Hawking-Ellis}:
\begin{eqnarray}
(\; \Delta  \; + \; m^{2} \;+\; 2\; ) \; A^{b} = 0  \;  , \;\; \xi
_{b}\; A^{b}  = 0 \; , \;\; L_{ab} A^{b}  = A_{a} \; , \nonumber
\\
dS^{2} = \;(1  - r^{2}) dt^{2} - { dr^{2} \over 1 - r^{2}} - r^{2}
( d \theta ^{2}  + \sin ^{2} \theta  d\phi ^{2}) \; \; .
\label{13.2.2}
\end{eqnarray}

\noindent Coordinates  $x^{\alpha } = (t, r, \theta , \phi  )$
and $\xi ^{a}$   are referred by
\begin{eqnarray}
\xi ^{1}  = r \sin \theta  \cos \phi  , \; \xi ^{2}  = r \sin
\theta  \sin \phi  , \; \xi^{3} = r \cos \theta  , \nonumber
\\
 \xi ^{0}  =
\sinh  \sqrt{1  - r^{2} }  , \; \xi ^{5}  = \cosh  \sqrt{1 -
r^{2}}   \; , \nonumber
\\
t  = \mbox{arctg}\; { \xi ^{0} \over \xi ^{5}} \; , \; r  = \sqrt{
(\xi ^{1})^{2}  + (\xi ^{2})^{2}  + (\xi ^{3})^{2} } \; ,
\nonumber
\\
\theta  = \mbox{arctg} \; { \sqrt{(\xi ^{1})^{2} + (\xi ^{2})^{2}}
 \over \xi ^{3}} \; , \;
\phi  = \mbox{arctg} \; { \xi ^{2} \over \xi ^{1}} \; .
\label{13.2.3}
\end{eqnarray}

\noindent These coordinates  $(t,\; r,\; \theta ,\; \phi  )$
cover the part of the full space \cite{1973-Hawking-Ellis}
\begin{eqnarray}
  \xi ^{5} \; + \; \xi ^{0} \ge  0 \; , \;
 \xi ^{5}  \; - \; \xi ^{0} \ge  0 \;.
 \nonumber
 \end{eqnarray}

For any representation of the group  $SO(4.1)$ on the functions
$\Psi (\xi )$, we have relationship
\begin{eqnarray}
\xi ' = S\; \xi \; , \;\; \Psi '(\xi ')  = U \; \Psi (\xi ) \qquad
\Longrightarrow \qquad  \Psi '(\xi )  = U\;  \Psi (S^{-1} \; \xi )
\; . \nonumber
\end{eqnarray}

\noindent In the case   $U \equiv  S$ and $\Psi  \equiv  A$, the
$(0-5)$-rotation
\begin{eqnarray}
\xi ^{0'}  = \cosh  \omega  \;  \xi ^{0}  +
             \sinh  \omega  \;  \xi ^{5} \; , \qquad
 \xi  ^{5'}  = \sinh \omega  \; \xi ^{0}  +
             \cosh  \omega  \;  \xi ^{5} \; ,
\nonumber
\end{eqnarray}

\noindent with an infinitesimal parameter $\delta \omega$ gives
\begin{eqnarray}
A' (\xi ) = ( I  \;+\; \delta \omega \;  J_{50} )\; A(\xi ) \; ,
\qquad J_{50}  =   L_{50}  + \sigma _{50}  \; , \nonumber
\\
 L_{50}  =  \xi _{5}\;{\partial  \over \partial \xi ^{0} } \;-\;
\xi ^{0} \; {\partial \over \partial \xi ^{5} }\; , \; \sigma
_{50} = \left | \begin{array}{ccccc}
0 & 0 & 0 & 0 & 1 \\
0 & 0 & 0 & 0 & 0 \\
0 & 0 & 0 & 0 & 0 \\
0 & 0 & 0 & 0 & 0 \\
1 & 0 & 0 & 0 & 0
\end{array} \right |   \; .
\label{13.2.6}
\end{eqnarray}

\noindent General expression for generators is (where  $g_{nb} =
\mbox{diag} (+1, -1,  -1, -1, -1)$)
\begin{eqnarray}
( J_{mn})^{a}_{\;\;b}  =  L_{mn} \; \delta ^{a}_{b} \;  + \;
(\sigma _{mn})^{a}_{\;\;b} \;\; , \;\;\; (\sigma
_{mn})^{a}_{\;\;b}  = \delta ^{\alpha }_{m}\; g_{nb}\;-\; \delta
^{a}_{n} \; g_{mb} \; . \label{13.2.7}
\end{eqnarray}

Let us search  solutions for eqs.  (\ref{13.2.2}) by dyagonalizing
three operators:
\begin{eqnarray}
( - i J_{50} )^{a}_{\;\;b} \; A^{b}  = \epsilon  A^{a} \; ,
\nonumber
\\
\qquad  ({\bf  J}^{2}) ^{a}_{\;\;b} A^{b} =
 j(j+1) A^{a} \; , \qquad ( J_{3})^{a}_{\;\;b}  A^{b}  =
 m  A^{a} \; ,
\label{13.2.8}
\end{eqnarray}

\noindent where
\begin{eqnarray}
J_{k} = - {i \over 2} \; \epsilon _{ijk} \; (  L_{ij} \;+\; \sigma
_{ij} ) =  l_{k}  \;+\; s_{k} \;  ,  \qquad  s_{1} = \left |
\begin{array}{ccccc}
 0 & 0 & 0 & 0 & 0 \\ 0
& 0 & 0 & 0 & 0 \\
0 & 0 & 0 & i & 0 \\
0 & 0 &-i & 0 & 0 \\
0 & 0 & 0 & 0 & 0     \end{array} \right | \; , \nonumber
\\
s_{2} =  \left | \begin{array}{ccccc}
0 & 0 & 0 & 0 & 0 \\ 0 & 0 & 0 & -i& 0 \\
0 & 0 & 0 & 0 & 0 \\ 0 & i & 0 & 0 & 0 \\ 0 & 0 & 0 & 0 & 0
\end{array} \right | \; ,  \qquad
s_{3} = \left | \begin{array}{ccccc}
0 & 0 & 0 & 0 & 0 \\
0 & 0 & i & 0 & 0 \\
0 &-i & 0 & 0 & 0 \\
0 & 0 & 0 & 0 & 0 \\
0 & 0 & 0 & 0 & 0     \end{array} \right | \; . \nonumber
\end{eqnarray}

First,  the eigenfunction equation $(- i J_{50})\; A = \epsilon
A$ is to be solved.  With the use of identity   $J_{50} =  -
\partial _{t} + \sigma _{50}$, we get
\begin{eqnarray}
{\bf  A} \sim e^{-i\epsilon t} \; , \qquad (A^{0} + A^{5}) \sim
e^{(-i\epsilon +1)t}\; , \qquad  (A^{0} - A^{5}) \sim
e^{(-i\epsilon -1)t}\; . \nonumber \label{13.2.9}
\end{eqnarray}

\noindent Bearing in mind two other equations in (\ref{13.2.8}),
for the 5-vector $A^{a}$  we  get the following substitution (see
\cite{Warshalovich})
\begin{eqnarray}
{\bf  A} = e^{-i\epsilon t} \; \left  [\; f(r) \; {\bf
Y}^{(j+1)}_{jm}\;+ \; g(r) \; {\bf  Y}^{(j-1)}_{jm} \; + \; h(r)
\;
 {\bf  Y} ^{(j)} _{jm} \; \right ] \; ,
\nonumber
\\
A^{0} = Y _{jm}   \; \left [ \;  e^{(-i\epsilon +1)t} F(r)   +
e^{(-i\epsilon -1)t} G(r) \; \right   ]\; , \nonumber
\\
A^{5} = Y_{jm}    \; \left [ \; e^{(-i\epsilon +1)t} F(r) -
e^{(-i\epsilon -1)t} G(r) \;  \right  ] \; . \label{13.2.10}
\end{eqnarray}

Radial functions $f(r),\; g(r), \;h(r), \; F(r), \; G(r)$  are to
be constructed on the base of eqs. (\ref{13.2.2}).  With the use
of the form of the operator $\Delta$ in variables   $(t,r,\theta
,\phi )$
\begin{eqnarray}
\Delta =   {1 \over 1 - r^{2}} {\partial ^{2} \over \partial t^{2}
} \;-\; {1 \over r^{2} } {\partial  \over \partial r} r^{2}(1 -
r^{2}) {\partial  \over \partial  r} \;-
 {1 \over r^{2}} ({1
\over \sin  \theta } {\partial  \over \partial  \theta } \sin
\theta {\partial  \over \partial  \theta } \;+\; {1 \over \sin
^{2}\theta } {\partial ^{2} \over \partial  \phi ^{2}} ) \nonumber
\label{13.2.11}
\end{eqnarray}

\noindent and  bearing in mind the  known known action of ${\bf
l} ^{2} $ on spherical functions  \cite{Warshalovich}
\begin{eqnarray}
{\bf  l} ^{2}  = - ( {1 \over \sin  \theta } {\partial  \over
\partial  \theta } \sin  \theta {\partial  \over \partial  \theta
} \;+\; {1 \over \sin ^{2}\theta } {\partial ^{2} \over \partial
\phi ^{2}} )\; , \nonumber
\\
{\bf  l}^{2} \;  {\bf  Y}^{(\nu )}_{jm} = \nu (\nu +1) {\bf
Y}^{(\nu )}_{jm}\;  ,  \;\; {\bf  l}^{2} \;  Y_{jm} = j (j +1)
Y_{jm} \; , \nonumber
\end{eqnarray}

\noindent for radial functions  $f(r), \;  g(r), \;  h(r), \;
F(r), \; G(r)$  we get  equations of  one the  same type
\begin{eqnarray}
\left [ {d^{2} \over d r^{2}} + {2(1 - 2 r^{2}) \over r(1 -
r^{2})} {d \over dr} - {\Lambda ^{2} \over (1 - r^{2})^{2} } \;-\;
{ m^{2} + 2 \over 1 - r^{2}} - { \nu (\nu  + 1) \over r^{2}(1 -
r^{2})}   \right  ]  U_{\Lambda ,\nu } = 0 \; ; \label{13.2.12a}
\end{eqnarray}

\noindent  radial functions are given by
\begin{eqnarray}
 f = f_{0} \; U_{-i\epsilon
,j +1}  \; ,  \qquad  g = g_{0}  U_{-i\epsilon , j-1}  \; , \qquad
h = h_{0}  U_{-i\epsilon ,j} \; , \nonumber
\\
F = F_{0}  U_{-i\epsilon +1,j} \; , \qquad G = G_{0} U_{-i\epsilon
+1, j}\; , \label{13.2.12b}
\end{eqnarray}

\noindent where $f_{0},\; g_{0}, \; h_{0}, \;F_{0},\;  G_{0}$ are
some constants. Solutions of  (\ref{13.2.12a})   can be expressed
in terms of hypergeometric  functions \cite{1953-Bateman-Erdelyi}
-- let us write down regular in  $r = 0$  ones (let $ z = r ^{2}=
\sin ^{2} \omega $):
\begin{eqnarray}
U_{-i\epsilon ,j}  = (\sin \omega )^{j}  \; (\cos  \omega
)^{-i\epsilon }\;  F(a , b , c ; z ) \; , \nonumber
\\
U_{-i\epsilon ,j+1} = (\sin  \omega )^{j+1} \; (\cos  \omega
)^{-i\epsilon } \; F(a + 1/2, b + 1/2, c + 1 ; z) \; , \nonumber
\\
U _{-i\epsilon ,j-1} = (\sin  \omega )^{j-1}\;
(\cos\omega)^{-i\epsilon} \; F(a - 1/2, b - 1/2, c - 1 ; z) \; ,
\nonumber
\\
U_{-i\epsilon +1,j} = (\sin \omega )^{j}\; (\cos  \omega
)^{-i\epsilon +1} \;  F(a + 1/2 , b  + 1/2, c ; z ) \; , \nonumber
\\
U_{-i\epsilon -1,j}  = (\sin \omega )^{j} \; (\cos  \omega
)^{-i\epsilon -1} \; F(a - 1/2 , b - 1/2, c ; z ) \; ;
\label{13.2.13a}
\end{eqnarray}

\noindent where
\begin{eqnarray}
a = { 3/2 + j + i \sqrt{m^{2}-1/4}  - i\epsilon  \over 2} \;  ,
\nonumber
\\
b = { 3/2 + j - i \sqrt{m^{2}-1/4}  - i\epsilon  \over 2} \;  ,
\;\;\;    c = j + 3/2\;  . \label{13.2.13b}
\end{eqnarray}

\noindent From additional constraint $\xi _{a} A^{a} = 0 $, with
the use of (see in \cite{Warshalovich})
\begin{eqnarray}
{\bf  \xi} \; {\bf  Y}^{(j)}_{jm} = 0 \;, \; {\bf  \xi}\; {\bf
Y}^{(j+1)}_{jm} = - \sqrt{{ j + 1 \over 2j + 1 }} \; r\; Y_{jm}
\; ,  \; {\bf  \xi} \; {\bf  Y}^{(j+1)}_{jm} = - \sqrt{{j \over
2j+1}} \; r \; Y_{jm}  \; , \nonumber
\end{eqnarray}

\noindent one gets
\begin{eqnarray}
- \sqrt{{ j + 1  \over 2j + 1}} \;  r\; f \;+\; \sqrt{{ j \over 2j
+ 1}} \; r \;g \;+  \sqrt{ 1 - r^{2} } \; ( G \;-\; F )   = 0 \; .
\label{13.2.14}
\end{eqnarray}

\noindent From the Lorentz condition  $L_{ab} \; A^{b}$ = $A_{a}$;
one gets (all details are omitted)
\begin{eqnarray}
- \sqrt{{j+1\over 2j  + 1 }}  ({d \over dr} + { j + 2 \over r} )\;
f \;+ \; \sqrt{{j \over 2j  + 1 }}  ({d \over dr} - { j - 1 \over
r} ) g  - \nonumber
\\
-
 {i \epsilon  \over \sqrt{1 - r^{2}}} (F + G)  = 0 \; ,
\label{13.2.18a}
\\
- \sqrt{{j + 1  \over 2j + 1 }}\; r \; f\; + \; \sqrt{{j \over 2j
+ 1 }} \; r \; g \;+\; \sqrt{1 - r^{2}} \; (F \;-\; G ) \; = 0 \;
; \label{13.2.18b}
\end{eqnarray}

\noindent the last equation coincides with  (\ref{13.2.14}). It is
convenient in expressions for  $f(r)$   and $g(r)$ to separate
special $j$-dependent factors
\begin{eqnarray}
\sqrt{{j + 1 \over 2j  + 1}}\; f(r)  = f_{0}\; U_{-i\epsilon ,j+1}
\; , \qquad  \sqrt{{j \over 2j  + 1}}\; g(r) = g_{0}\;
U_{-i\epsilon ,j-1} \; .
 \label{13.2.20}
 \end{eqnarray}

\noindent Bearing in mind  (\ref{13.2.20}),  eqs. (\ref{13.2.18a})
-- (\ref{13.2.18b}) are changed to
\begin{eqnarray}
f_{0} ( - {d \over d \omega  } + {j+2 \over \tan  \omega  } )\;
U_{-i\epsilon , j+1} + g_{0}  ( {d \over d \omega  } - {j-1 \over
\tan
 \omega  } )\;
U_{-i\epsilon ,j-1}  - \nonumber
\\
\nonumber - i \epsilon  (\; F_{0}  U_{-i\epsilon +1, j}\; +\;
G_{0}
 U_{-i\epsilon -1, j} \; )   = 0 \; ,
\label{13.2.21a}
\\
 - f_{0} \; \tan  \omega  \; U_{-i\epsilon ,j+1} \;+ \; g_{0}
\; \tan \omega \; U_{-i\epsilon ,j-1} \;+ \nonumber
\\
+ \; F_{0} \; U_{-i\epsilon +1,j} \; -\;  G_{0} U_{-i\epsilon
-1,j} = 0 \; . \label{13.2.21b}
\end{eqnarray}

\noindent These relations may be resolved as follows
\begin{eqnarray}
2\; F_{0}  U_{-i\epsilon +1,j} = f_{0}   [ - {1 \over i\epsilon }
( - {d \over d \omega  } + {j+2 \over \tan  \omega } ) + \tan
\omega   ]  U_{-i \epsilon ,j+1} + \nonumber
\\
+   g_{0}   [  + {1 \over i\epsilon }   ( {d \over d \omega } -
{j-1 \over \tan  \omega  } )  -   \tan  \omega    ] U_{-i\epsilon
,j-1}  \; , \nonumber \label{13.2.22a}
\\[2mm]
2 G_{0} U_{-i\epsilon -1,j} =
 f_{0}   [  -{1 \over i\epsilon }
( - {d \over d \omega } + {j+2 \over \tan \omega} ) -  \tan \omega
]  U_{-i\epsilon ,j+1}  + \nonumber
\\
+ g_{0}   [  +{1 \over i\epsilon }
  (  {d \over d \omega  } - {j-1 \over \tan \omega }   )
+ \tan \omega   ]  U_{-i\epsilon ,j-1}   \; . \label{13.2.22b}
\end{eqnarray}

\noindent From whence, bearing in mind expressions for
$U_{-i\epsilon \pm 1,j}$ and  $U_{-i\epsilon ,j\pm 1}$  in terms
of hypergeometric functions, we arrive at
\begin{eqnarray}
2  F_{0} F(a + 1/2 , b  + 1/2 , c ; z )  =
 f_{0}   [- { 2j+3 \over i \epsilon  } F(a + 1/2 , b + 1/2 ,
c + 1 ; z )- \nonumber
\\
-  {2 z \over i \epsilon } {d \over d z} F(a + 1/2 , b + 1/2 , c +
1 ; z )  ]  +
   g_{0}  { 2 \over i \epsilon } {d \over d z}
F(a-1/2,b-1/2,c-1;z) , \nonumber
\\
\label{13.2.23a}
\end{eqnarray}
\begin{eqnarray}
 2\; G_{0}   F(a -1/2 , b - 1/2 , c ; z ) =
 f_{0} (1 - z)   [ - {2j+3 \over i \epsilon }
 F(a + 1/2, b + 1/2, + 1 ; z )  -
\nonumber
\\
 -   { 2z \over 1 -z } F(a + 1/2 , b + 1/2 , c +
1 ; z ) - {2 z \over i \epsilon } {d \over d z} F(a + 1/2 , b  +
1/2 , c + 1 ; z )  ] \;+ \nonumber
\\
+ g_{0}  [   2 F( a - 1/2 , b - 1/2 , c - 1 ; z )  + {2 z \over i
\epsilon } (1 -  z)  {d \over d z} F (a - 1/2 , b - 1/2 , c - 1 ;
z )   ]   . \nonumber
\\
\label{13.2.23b}
\end{eqnarray}

\noindent By simplicity reason, let us search for solutions of the
types:
\begin{eqnarray}
(j+1)- \mbox{wave} \; , \qquad f_{0} \neq  0  \; , \; g_{0} = 0 \;
,\; h_{0} = 0 \; ; \nonumber
\\
(j-1)-\mbox{wave} \; , \qquad f_{0} = 0 \; , \; g_{0} \neq  0 \;
,\; h_{0} = 0 \; ; \nonumber
\\
j -\mbox{wave} \; , \qquad   f_{0} = 0 \; , \; g_{0} = 0 \; ,\;
h_{0} \neq  0 \; . \label{13.2.24}
\end{eqnarray}

\noindent The task is to satisfy  (\ref{13.2.23a}) --
(\ref{13.2.23b}) and determine the $F_{0}$  and $G_{0}$ in
dependence of  three factors $f_{0} , g_{0} , h_{0}$ . For
$j$-wave, from (\ref{13.2.23a})  --  (\ref{13.2.23b})  it follows
 $F_{0} = 0$  and $G_{0} = 0$ ;
that is  in this case the components  $A^{0}$  and  $A^{5}$
vanish. In the case of  $(j-1)$-wave, relation   (\ref{13.2.23a})
takes the form
\begin{eqnarray}
2 \; F_{0}\; F(a  + 1/2 , b + 1/2 , c ; z )  = g_{0} \; {2 z \over
i \epsilon }\; {d \over d z} \; F( a - 1/2 , b -1/2 , c - 1 ; z
)\; ; \nonumber
\end{eqnarray}

\noindent from  whence  with the use of the formula
\cite{1953-Bateman-Erdelyi}
\begin{eqnarray}
{d \over d z} \; F(\alpha , \beta , \gamma ; z ) = { \alpha \beta
\over \gamma } \; F(\alpha  + 1 , \beta  + 1 , \gamma  + 1; z)
\nonumber
\end{eqnarray}

\noindent  we immediately produce
\begin{eqnarray}
(j-1)-\mbox{wave}\; , \qquad  F_{0} =   \; { (a - 1/2) \; (b -
1/2) \over i \epsilon \; (c - 1)}  \; . \label{13.2.25a}
\end{eqnarray}

\noindent Eq.  (\ref{13.2.23b}) for  $(j-1)$-wave  gives
\begin{eqnarray}
2 G_{0} \; F(a -1/2 , b - 1/2 , c ; z ) = 2 \; g_{0} \; [\; F (a -
1/2, b - 1/2, c - 1 ; z ) \; + \nonumber
\\
+ \; {1 - z \over i \epsilon  }\; {d \over d z}\; F ( a - 1/2 , b
- 1/2 , c - 1 ; z )\; ] \; , \nonumber
\end{eqnarray}

\noindent and further  with the use of the known  relation
\cite{1953-Bateman-Erdelyi}
\begin{eqnarray}
{d \over d z}\; F(\alpha  , \beta  , \gamma  ; z )  =
 {\alpha  + \beta  + \gamma  \over 1  - z } \;
F(\alpha  , \beta  , \gamma  ; z )  \; + \; {(\alpha  - \gamma )\;
(\beta - \gamma ) \over \gamma ( 1  - z)} \; F( \alpha  , \beta  ,
\gamma  + 1 ; z) \nonumber
\end{eqnarray}

\noindent we get the factor  $G_{0}$
\begin{eqnarray}
(j-1)-\mbox{wave} \;  , \qquad  G_{0} = {(a - c + 1/2) \; (b - c +
1/2) \over i \epsilon (c - 1) } \; g_{0}  \; . \label{13.2.25}
\end{eqnarray}

For  $(j  +  1)$-wave, relation (\ref{13.2.23a} gives
\begin{eqnarray}
2\; F_{0} \; F(a + 1/2 , b  + 1/2 , c ; z )  = f_{0} \;  [ - {
2j+3  \over i \epsilon } \; F(a + 1/2 , b + 1/2 , c + 1 ; z ) \; -
\nonumber
\\
- \; {2 z \over i \epsilon } \; {d \over d z} \; F(a + 1/2 , b +
1/2 ,  c +1; z ) \; ]    \; , \nonumber
\end{eqnarray}

\noindent and further with the use of the formula
\cite{1953-Bateman-Erdelyi}
\begin{eqnarray}
z\; {d \over d z}\; F(\alpha  , \beta  , \gamma  ; z ) = (\gamma -
1)\;  [ \; F(\alpha  , \beta  , \gamma  - 1 ; z )\; - \; F(\alpha
, \beta  , \gamma  ; z ) \; ] \; , \nonumber \label{13.2.26a}
\end{eqnarray}

\noindent we get the factor $F_{0}$
\begin{eqnarray}
(j + 1)- \mbox{wave} \; ,\qquad  F_{0} =  { c \over - i
\epsilon}\;f_{0}\; . \label{13.2.26b}
\end{eqnarray}

\noindent in turn, relation  (\ref{13.2.23b}) takes the form
\begin{eqnarray}
2  G_{0}  F(a -1/2 , b - 1/2 , c ; z ) = f_{0} (1 - z)  [ - { 2j+3
\over i \epsilon }  F(a + 1/2 , b + 1/2 , c + 1 ; z) - \nonumber
\\
-   {2 z \over 1 - z }  F(a + 1/2 , b  + 1/2 , c  + 1 ; z ) -
  {2 z \over i \epsilon} {d \over d z}
 F(a + 1/2 , b  + 1/2 , c + 1 ; z )  ] \; ,
\nonumber \label{13.2.27a}
\end{eqnarray}

\noindent which with the use of  (\ref{13.2.26a}) gives
\begin{eqnarray}
- i \epsilon  \; { G_{0} \over f_{0} }  \;  F(a - 1/2 , b - 1/2 ,
c ; z ) = \nonumber
\\
= (c - a - b)  z  F(a + 1/2 , b + 1/2 , c + 1 ; z )  + c (1 - z)
F(a + 1/2 , b + 1/2 , c ; z )  \; . \nonumber
\end{eqnarray}

\noindent Let us differentiate the  later:
\begin{eqnarray}
- i \epsilon  \; { G_{0} \over f_{0} } \; {(a - 1/2)(b - 1/2)
\over c } F(a + 1/2, b + 1/2, c + 1 ; z )  = \nonumber
\\
=( c - a - b) \;  ( z {d \over d z } + 1  ) \;  F(a + 1/2, b +
1/2, c + 1 ; z ) \; + \; \nonumber
\\
c \; ( - 1 \;+\; ({1 -z \over z}) \; {d \over d z }\; ) \; F(a +
1/2, b + 1/2, c ; z ) \;  , \nonumber
\end{eqnarray}

\noindent from whence  we get
\begin{eqnarray}
- i \epsilon  \; { G_{0} \over f_{0} } \; {(a - 1/2) (b - 1/2)
\over c } \; F(a  + 1/2, b  + 1/2, c  + 1 ; z ) = \nonumber
\\
= c ( - {c - 1 \over z} + 2 c - a - b - 2)  ) \; F(a + 1/2,b +
1/2, c; z ) \; + \nonumber
\\
+ c (c - 1)\; {1 - z \over z}\; F(a + 1/2 , b + 1/2 , c - 1 ; z )
\; - \nonumber
\\
-(1 - c)(c - a - b) \; F (a + 1/2 , b + 1/2 , c ; z) \;  .
\label{13.2.27b}
\end{eqnarray}

\noindent Not, let us use the known identity for hypergeometric
functions \cite{1953-Bateman-Erdelyi}
\begin{eqnarray}
c    [ (c - 1) - (2c - a - b - 2) z   ] F (a + 1/2 , b + 1/2 , c ;
z)  + \nonumber
\\
+  (c - a - 1/2)  (c - b - 1/2)  z  F(a + 1/2, b + 1/2, c + 1 ; z)
- \nonumber
\\
-  c (c - 1) (1 - z)
 F(a + 1/2 , b + 1/2 , c - 1 ; z )   = 0 \; ,
\nonumber
\end{eqnarray}

\noindent then (\ref{13.2.27b}) results in
\begin{eqnarray}
- i \epsilon   { G_{0} \over f_{0} }  {(a- 1/2)(b - 1/2) \over c }
F(a + 1/2, b + 1/2, c + 1 ; z ) = \nonumber
\\
= [ (c - a - 1/2) (c - b - 1/2) + (1 - c) (c - a - b) ] F(a + 1/2,
b + 1/2, c + 1 ; z )  \;  ; \nonumber
\end{eqnarray}

\noindent therefore the factor $G_{0}$ is given by
\begin{eqnarray}
(j + 1)-\mbox{wave} \; , \qquad  G_{0} = { c \over  - i \epsilon
} \; f_{0} \;\; . \label{13.2.27c}
\end{eqnarray}

Collecting the results obtained

$\nu  = j \; ,$
\begin{eqnarray}
  {\bf  A} = e^{-i\epsilon t} h_{0} \;
U_{-i\epsilon ,j} \; {\bf  Y}^{(j)}_{jm}(\theta ,\phi ) \; , \;
A^{0} = 0  \; , \; A^{5} =  0  \;     ; \label{13.2.28a}
\end{eqnarray}
$\nu  = j+1 $ ,
\begin{eqnarray}
  {\bf  A}  = e^{-i\epsilon t}
{\bf  Y}^{(j+1)} _{jm} \; \sqrt{{2j + 1 \over j + 1 }} \; , \;
F_{0} = -  { c \over i \epsilon } f_{0} \; , \; G_{0}  = - { c
\over i \epsilon } f_{0} \;       ; \label{13.2.28b}
\end{eqnarray}
$\nu  = j-1 $,
\begin{eqnarray}
 {\bf  A}  = e^{-i\epsilon t}
{\bf  Y}^{(j-1)}_{jm} \; \sqrt{{2j + 1 \over j}} \; g_{0} \;
U_{-i\epsilon ,j -1} \; , \;\; \nonumber
\\
F_{0}  = { (a - 1/2) (b - 1/2) \over i \epsilon  (c - 1)} g_{0}\;
, \;\; G_{0}  = { (a - c + 1/2) (b - c + 1/2)  \over i \epsilon (c
- 1) } g_{0} \; ; \label{13.2.28c}
\end{eqnarray}

\noindent  in the cases  (\ref{13.2.28b}) and  (\ref{13.2.28c})
the components  $A^{0}$  and $A^{5}$ are to be calculated by the
same formulas with different values of $F_{0}$  and $G_{0}$:
\begin{eqnarray}
A^{0} = Y_{jm}  \; \left [ \;  e^{(-i\epsilon +1)t}\;  F_{0}\;
U_{-i\epsilon +1,j}  \;+\; e^{(-i\epsilon -1)t} \; G_{0} \;
U_{-i\epsilon -1,j}  \;  \right ]  \; , \nonumber
\\
A^{5} = Y_{jm} \;  \left [ \;  e^{(-i\epsilon +1)t} \; F_{0} \;
U_{-i\epsilon +1,j} - e^{(-i\epsilon -1)t} \; G_{0} \;
U_{-i\epsilon -1,j}  \; \right ] \; . \label{13.2.28d}
\end{eqnarray}

\section{General method to construct electromagnetic  $\Pi,E,M$-waves}

Let us recall the situation in the flat Minkowski space. If a
scalar function $\Lambda (x)$ the massless  wave equation
\begin{eqnarray}
\Delta \; \Lambda  (x) = 0  \; ,\; \Delta  =  \partial ^{0}
\partial _{0}\; -\; \partial ^{i} \partial _{i} \; ,
\nonumber \label{1.1}
\end{eqnarray}

\noindent then three linearly independent solutions of the  vector
wave equation $\Delta  \; A^{\alpha} (x) =  0$  can be constructed
as follows \cite{1948-Stratton}:
\begin{eqnarray}
A^{(1)}_{\alpha } = {\partial  \over \partial  x^{\alpha } }
\Lambda (x) \; ,\qquad   {\bf  A}^{(2)} = {\bf  r} \times  {\bf
A}^{(1)} \; , \qquad {\bf  A}^{(3)}  = \nabla  \times  {\bf
A}^{(2)} \; . \label{1.2}
\end{eqnarray}

\noindent If the  $\Lambda (x)$ is taken as a spherical wave
\begin{eqnarray}
\Lambda (x) = e^{-i\epsilon t} \; Y_{jm}(\theta ,\phi ) \;
f_{j}(\epsilon r) \; , \nonumber \label{1.3}
\end{eqnarray}

\noindent $f_{j}(\epsilon r)$ is a Bessel spherical function, the
recipe (\ref{1.2}) gives three spherical vector solutions
\cite{1948-Stratton}:
\begin{eqnarray}
{\bf  A}^{(1)} \sim e^{-i\epsilon t} \; [\; \sqrt{{j \over 2 j +
1}} \; f_{j-1} \; {\bf  Y}^{(j-1)}_{jm} \; +  \; \sqrt{{j + 1
\over 2 j + 1 }}\; f_{j+1} \; {\bf  Y}^{(j+1)}_{jm}\; ]\; ,
\nonumber
 \\
 \;\; A^{(1)0} \sim  i \epsilon  e^{-i\epsilon t}\; Y_{jm} f_{j} \;, \qquad \partial ^{\alpha } A^{(\Pi)}_{\alpha } = 0  ;
\nonumber
\\
{\bf  A}^{(2)} \sim e^{-i\epsilon t} \; f_{j}(\epsilon r) \; {\bf
Y}^{(j)}_{jm}(\theta ,\phi ) \; , \qquad A^{(2)0} = 0 \; , \qquad
\mbox{div} \; {\bf  A}^{(M)} = 0 \; ; \nonumber
\\
{\bf  A}^{(3)} \sim e^{-i\epsilon t}\;  [\; \sqrt{{j + 1 \over 2 j
+ 1}} \; f_{j-1}  \; {\bf  Y}^{(j-1)}_{jm}\; - \; \sqrt{{j \over 2
j + 1 }} \; f_{j+1}\; {\bf  Y}^{(j+1)}_{jm} \; ]\; , \nonumber
\\
 A^{(3)0} = 0 \; , \qquad \mbox{div} \; {\bf  A}^{(E)} = 0 \; ,
\label{1.4c}
\end{eqnarray}

\noindent which are called respectively   $\Pi-, M-, E-$waves. The
task is to extend this method to de Sitter space starting with
5-form:
\begin{eqnarray}
( \Delta  + 2 ) A^{b} = 0\; ,\;  \Delta  = - {1 \over 2}  L^{ab}
L_{ab} = m^{a} m_{a}\; ,\; L_{ab}  A^{b} = 0 \; , \;  \xi ^{a}
A_{a} = 0 \; . \nonumber \label{1.6}
\end{eqnarray}

\noindent Evidently, that $\Pi$-wave should be determined by the
rule
\begin{eqnarray}
A^{(\Pi)}_{a} = m_{a} \; \Lambda (x) \; ,\;\; \Delta \; \Lambda
(x) = 0 \; . \label{1.7}
\end{eqnarray}

\noindent Let us demonstrate with the help of commutative
relations  that  the 5-vector  $m_{a}  \Lambda (x)$  satisfies $(
\Delta  + 2 ) A^{b} = 0 $. Indeed, bearing in mind $ [ m_{a}  ,
m_{b}  ] = L_{ab} , $ one may obtain
\begin{eqnarray}
\Delta  m_{a}  \Lambda (x) = [  m^{b}  m_{b}  ,  m_{a} ]  \Lambda
(x) = ( L_{ab}  m^{b}  +  m^{b}  L_{ab}) \Lambda (x)\; ; \nonumber
\end{eqnarray}

\noindent which with the help of $[  m^{b} ,  L_{ab} ]  = - 4
m_{a}\; $ reduces to
\begin{eqnarray}
\Delta \; m_{a} \; \Lambda (x) = (\; 2 L_{ab}\; m^{b} \; + \; [\;
m^{b}\; , \; L_{ab}\; ]\; )\; \Lambda (x)  = (\; 2 L_{ab} \; m^{b}
\; - \; 4 m_{a}\; )\; \Lambda (x)\; , \nonumber
\end{eqnarray}

\noindent from whence it follows
\begin{eqnarray}
( \Delta  +  2  )  m_{b}  \Lambda (x) = 0 \; . \label{1.11}
\end{eqnarray}

Let us define $M$-wave. One might expect the structure
\begin{eqnarray}
{\bf  A}^{(M)} = \vec{\xi } \times  {\bf  A}^{(\Pi)} = ( \vec{\xi
} \times  {\bf  m} ) \; \Lambda (x) \; , \qquad \Delta  \Lambda
(x) =  0 \; . \label{1.12}
\end{eqnarray}

\noindent Because the operator $ ( \vec{\xi } \times  {\bf  m}
)_{i} = \epsilon _{ijk} \xi _{j}  ( \partial _{k}  +  \xi _{k}
\xi ^{a} \partial _{a} ) $ commutes with $\Delta $, the equation $
\Delta   ( \vec{\xi } \times  {\bf  m} ) \Lambda (x) = 0 \; ; $
holds, from whence  we conclude that  one must make   the starting
structure (\ref{1.12}) more exact
\begin{eqnarray}
{\bf  A}^{(M)} = (\vec{\xi } \times  {\bf  m} )  K(x) \; , \; (
\Delta  +  2 )    K(x) = 0 \; ; \label{1.13}
\end{eqnarray}

\noindent where the scalar function  $K(x)$  satisfies the
conformally invariant equation. It remain to prove the  structure
for $E$-wave:
\begin{eqnarray}
{\bf  A}^{(E)} = {\bf  m} \times  (\; \vec{\xi } \times {\bf  m}
\; )\;
 \Lambda (x) \; .
\label{1.14}
\end{eqnarray}

\noindent Indeed, the relationship
\begin{eqnarray}
( \Delta  + 2 )  A^{(E)} _{n}  = (  \Delta  +  2  ) \epsilon
_{nij}   m_{i}  \epsilon _{jkl}  \xi _{k}   m_{l}  \Lambda (x) =
{1  \over 2}  \epsilon _{nij}  \epsilon _{jkl} (  \Delta  +  2 )
m_{i}  L_{kl}  \Lambda (x) \; ; \nonumber
\end{eqnarray}

\noindent  with the help of identity $ [  m_{i} , L_{kl}  ]  = (
g_{ik}  m_{l}  - g_{il}  m_{k} ), $ transforms to
\begin{eqnarray}
( \Delta +  2 ) A^{(E)}_{n}  = {1 \over 2} \epsilon _{nij}
\epsilon _{jkl}  ( \Delta   +  2 ) ( L_{kl}  m_{i}  +  g_{ik}
m_{l}  -  g_{il}  m_{k}  )
 \Lambda (x)  =
\nonumber
\\
{1 \over 2} \epsilon _{nij}  \epsilon _{jkl} [ L_{kl} ( \Delta  +
2 ) m_{i}   + g_{ik} (\Delta  + 2 ) m_{l} - g_{il} (\Delta  +  2 )
m_{k} ] \Lambda (x) \; ; \nonumber
\end{eqnarray}

\noindent from this, taking into account $(\Delta \; +\; 2 )\;
m_{b} \; \Lambda (x) = 0 $ ,  one arrives at
\begin{eqnarray}
(\; \Delta \; + \; 2 \;) \; A^{(E)} _{n} =  0 \;  . \label{1.15}
\end{eqnarray}

\noindent Thus,  solutions of the types $\Pi-,M-,  E-$  in de
Sitter space are determined by
\begin{eqnarray}
A^{(\Pi)}_{a}  = m_{a} \; \Lambda (x) \; , \qquad \Delta  \;
\Lambda (x)  = 0 \;\;  ; \nonumber
\\
{\bf  A}^{(M)} = (\vec{\xi} \times  {\bf  m} ) \; K(x) \; , \qquad
(\; \Delta \; +   \; 2 \;) \; K(x)  = 0  \;\; ; \nonumber
\\
{\bf  A}^{(E)} = {\bf  m} \times  (\; \vec{\xi } \times  {\bf  m}
\;)\; \Lambda (x)\; , \; \Delta \; \Lambda (x) = 0 \; .
\label{1.16}
\end{eqnarray}

\noindent To obtain  $A^{0}$  and  $A^{5}$ for  $M -,E -$waves one
should use equations (see (\ref{1.6}))
\begin{eqnarray}
L_{ab} \; A^{b} = 0 \; , \qquad  \xi ^{a} \; A_{a} = 0 \; .
\nonumber
\end{eqnarray}

\section{Spherical  $E,M,\Pi$-waves in static coordinates
}

Starting with static spherical coordinates in de Sitter space
\begin{eqnarray}
dS^{2} =  (1 - r^{2}) dt^{2} \;-\; { dr^{2} \over 1 - r^{2}} \;-\;
r^{2} \; ( d\theta ^{2} + \sin ^{2} \theta  d\phi ^{2} ) \;
 ,
\label{2.1}
\end{eqnarray}

\noindent a scalar solution for $\Delta \; \Lambda (x) = 0$  let
us choose  as follows
\begin{eqnarray}
\Delta \; \Lambda (x) = 0 \;    ,\; \Lambda (x) = e^{-i\epsilon t}
\; Y_{jm}(\theta ,\phi ) \; f(r) \;  , \nonumber
\\
f(r) = \sin ^{j} \omega \; (\cos \omega )^{-i\epsilon}\; F (a -
1/2, b + 1/2, c ; z )\; , \nonumber
\\
a =  {j + 1 - i \epsilon  \over 2} \; ,\;\; b = {j + 2 - i
\epsilon  \over 2} \; , \;\;  c = j + 3/2 \; . \label{2.2}
\end{eqnarray}

\noindent We  need explicit form for  $m_{b}$  in  these
coordinates $(t,r,\theta ,\phi )$
\begin{eqnarray}
(\; m^{0} \;+\; m^{5}\;)  = e^{+t}\;  [ \; {1 \over \sqrt{1 -
r^{2}}} \;{\partial \over \partial t} \;+\; r\; \sqrt{1 - r^{2}}\;
{\partial \over \partial r} \; ]\;, \nonumber
\\
(\; m^{0} \;-\; m^{5}\; ) = e^{-t} \;  [\; {1\over \sqrt{1 -
r^{2}}}\;{\partial \over \partial t}\;-\; r\;\sqrt{1 -
r^{2}}\;{\partial  \over \partial r} \;  ] \; , \nonumber
\\
( m^{i})  = {\bf  m} = (\; - \nabla \; +\; \vec{\xi }\; r
{\partial  \over \partial r}\; ) \; , \;\; \nabla = {\partial
\over \partial  \vec{\xi } }  \; . \label{2.3}
\end{eqnarray}

With the use of known properties of spherical functions
\cite{Warshalovich},   we get representation for $\Pi$-wave
\begin{eqnarray}
\Pi -\mbox{wave} \; , \qquad ( A^{0}\; + \; A^{5})^{\Pi}  =
e^{(-i\epsilon +1)t}\; Y_{jm} \; (\; - { i \epsilon  \over \cos
\omega } \;+\; \sin  \omega {d \over d\omega } \;  )\; f\; ,
\nonumber
\\
( A^{0} \;-\; A^{5})^{\Pi} = e^{(-i\epsilon -1)t} \; Y_{jm}\; ( \;
- {i \epsilon  \over \cos \omega} \;-\; \sin  \omega {d \over
d\omega } \;  ) \; f\; , \nonumber
\\
{\bf A}^{\Pi} = e^{-i\epsilon t}\;  [ \; \sqrt{{j + 1  \over 2j +
1}} \; {\bf  Y}^{(j+1)}_{jm} \;  ( \cos  \omega  {d \over d\omega
} \;-\; {j \over \sin  \omega }  )\; f\; -\; \nonumber
\\
- \; {\bf  Y}^{(j-1)}_{jm}\;  ( \cos  \omega  {d \over d\omega }
\;+\; {j+1 \over \sin  \omega }  )\; f \;  ] \; . \label{2.4}
\end{eqnarray}

\noindent For $M$-wave,  using identities \cite{Warshalovich}
 \begin{eqnarray}
\vec{\xi } \times  {\bf  Y}^{(j+1)}_{jm}  = i \; \sqrt{{ j \over
2j + 1 }}\; r\; {\bf  Y}^{(j)}_{jm}\; , \qquad \vec{\xi } \times
{\bf  Y}^{(j-1)}_{jm}  = i\; \sqrt{{ j+1 \over 2j + 1 }}\;
 r\; {\bf  Y}^{(j)}_{jm}\; ,
\nonumber
\end{eqnarray}

\noindent we get
\begin{eqnarray}
{\bf  A}^{(M)} = {i \over \sqrt{j(j + 1)}}\; (\; \vec{\xi } \times
{\bf  m}\; ) \; K(x) = e^{-i\epsilon t} \; {\bf  Y}^{(j)}_{jm}
(\theta ,\phi ) \; U_{-i\epsilon ,j}(z) \; ; \label{2.5}
\end{eqnarray}

\noindent where  $U_{-i\epsilon ,j}(z)$  is determined by
\begin{eqnarray}
(\Delta \; +\; 2 ) \; K(x)  = 0 \; ,\;\; K(x) = e^{-i\epsilon t}\;
Y_{jm}(\theta ,\phi ) \; U_{-i\epsilon ,j}(z) \; , \nonumber
\\
U_{-i\epsilon ,j}(z)  = \sin ^{j}\omega \; (\cos \omega
)^{-i\epsilon } \; F ( a , b , c ; z ) \; . \label{2.6}
\end{eqnarray}

\noindent For $E$-wave we get
\begin{eqnarray}
{\bf  A}^{(E)} = {i \over \sqrt{j(j + 1)}}\;
 [\; {\bf  m} \times  (\; \vec{\xi } \times  {\bf  m}\; )\; ]\;
  \Lambda (x) = {\bf  m}  \; \times  [\; e^{-i\epsilon t}\;
{\bf  Y}^{(j)}_{jm} (\theta ,\phi ) \; f(r) \;]\; ,
\nonumber
\end{eqnarray}

\noindent and further
\begin{eqnarray}
{\bf  A}^{(E)} = e^{-i\epsilon t} \;  \left  [ \; \sqrt{{j \over
2j + 1}} \; {\bf  Y}^{(j+1)}_{jm} \; ( \cos  \omega  {d \over
d\omega } \;-\; {j \over \sin  \omega }  ) \; f \; + \right.
\nonumber
\\
\left. \; \sqrt{{j+1 \over 2j + 1}} \; {\bf  Y}^{(j-1)}_{jm} \; (
\cos \omega  {d \over d\omega } \;+\; {j+1 \over \sin  \omega }
)\; f \; \right  ]   \; . \label{2.7b}
\end{eqnarray}

Supposing that $(\Pi, E)$-waves are linear combinations of
solutions $(j+1) $  and  $(j-1)$ constructed in Section {\bf 2}
(at $m=0$),  we reduce the task to studying four relations
\begin{eqnarray}
( - { i \epsilon  \over \cos  \omega } \;+\; \sin  \omega  {d
\over dr}  ) \; f = \mbox{const} \; U_{-i\epsilon +1,j} \;  ,
\label{2.8a}
\\
( - { i \epsilon  \over \cos  \omega } \;-\; \sin  \omega  {d
\over dr}  ) \; f = \mbox{const} \; U_{-i\epsilon -1,j} \; ,
\label{2.8b}
\\
 ( \cos  \omega  {d \over d\omega } \;-\; {j \over \sin  \omega }
) \; f = \mbox{const} \; U_{-i\epsilon ,j+1} \; ,
 \label{2.8c}
\\
( \cos  \omega  {d \over d\omega } \;+\; {j+1 \over \sin  \omega }
) \; f = \mbox{const}  \; U_{-i\epsilon ,j-1} \; ; \label{2.8d}
\end{eqnarray}

\noindent the four constants are to be found. In the case
(\ref{2.8a})  we  have
\begin{eqnarray}
( - { i \epsilon  \over \cos  \omega } \;+\; \sin  \omega  {d
\over dr}  ) \; f = \nonumber
\\
= \sin ^{j}\omega \; (\cos \omega )^{-i\epsilon +1}\;  [ \; (j - i
\epsilon ) \; F(a - 1/2, b + 1/2, c ; z ) \;+ \nonumber
\\
+  \; 2 z {d \over dz}\; F (a - 1/2, b + 1/2, c ; z) \;]\; ,
\nonumber
\end{eqnarray}

\noindent from whence, using the rule  \cite{1953-Bateman-Erdelyi}
\begin{eqnarray}
z {d \over dz} \; F(\alpha , \beta , \gamma ; z) = \alpha \;  [\;
F(\alpha  + 1, \beta , \gamma  ; z) \;-\; F( \alpha  , \beta ,
\gamma  ; z) \; ] \; , \nonumber
\end{eqnarray}

\noindent we get
\begin{eqnarray}
( - { i \epsilon  \over \cos  \omega } \;+\; \sin  \omega  {d
\over dr}  ) \; f = \nonumber
\\
= \sin ^{j}\omega  (\cos \omega )^{-i\epsilon +1}  (j - i \epsilon
)  F(a + 1/2, b + 1/2, c ; z )  = (j - i \epsilon ) U_{-i\epsilon
+1,j}  . \label{2.9}
\end{eqnarray}

\noindent In the case  (\ref{2.8b})
\begin{eqnarray}
 ( - { i \epsilon  \over \cos  \omega }  -
\sin  \omega  {d \over dr}  )  f = \nonumber
\\
= \sin ^{j}\omega\; (\cos \omega )^{-i\epsilon -1} [ - ( j - i
\epsilon  )  F + ( j + i \epsilon  )  z  F - 2 z   {d \over dz}  F
] = \nonumber
\\
= \sin ^{j}\omega  (\cos \omega )^{-i\epsilon -1}   [ - 2 i
\epsilon \; F(a - 1/2, b + 1/2, c ; z)  - \nonumber
\\
- 2 (a - 1/2)  (1 - z)  F (a + 1/2 , b + 1/2 , c ; z )  ]\;  ;
\nonumber
\end{eqnarray}

\noindent and further, with the use of the known identity for
hypergeometric functions \cite{1953-Bateman-Erdelyi} we get
\begin{eqnarray}
[ (\gamma - \alpha  - \beta ) \; F ( \alpha  , \beta  , \gamma  ;
z) \; + \nonumber
\\
+ \; \alpha \; (1 - z)\; F(\alpha  + 1, \beta , \gamma  ; z) \;-\;
(\gamma  - \beta ) \; F(\alpha , \beta  - 1, \gamma  ; z)\; ]   =
0 \nonumber
\end{eqnarray}

\noindent (at $\alpha = a - 1/2 , \beta  = b + 1/2 , \gamma  = c
)$; therefore
\begin{eqnarray}
( - { i \epsilon  \over \cos  \omega }  + \sin  \omega  {d \over
dr}  )  f = \nonumber
\\
= - (j + i \epsilon )  \sin ^{j}\omega  (\cos \omega )^{-i\epsilon
+1} F (a - 1/2 , b - 1/2 , c ; z ) = - (j + i \epsilon )
U_{-i\epsilon -1,j} . \nonumber
\\
\label{2.10}
\end{eqnarray}

\noindent Let us consider eq.  (\ref{2.8c}):
\begin{eqnarray}
( \cos  \omega  {d \over d\omega } \;-\; {j \over \sin  \omega }
) \; f = \sin ^{j+1}\omega \; (\cos \omega )^{-i\epsilon } [ (- j
+ i \epsilon )\; F  \;+\; 2 (1 - z) \;{d\over dz}\; F\; ]\;  ;
\nonumber
\end{eqnarray}

\noindent allowing for the rule  \cite{1953-Bateman-Erdelyi}
\begin{eqnarray}
(1 - z) \; {d\over dz}\; F(\alpha , \beta , \gamma  ; z)  = [\;
(\alpha + \beta - \gamma ) \; F(\alpha , \beta , \gamma  ; z)
\;+\; \nonumber
\\
+ \; {(\alpha  - \gamma ) \; (\beta  - \gamma ) \over \gamma  }\;
F ( \alpha , \beta  , \gamma  + 1 ; z ) \;]\; , \nonumber
\end{eqnarray}

\noindent we arrive at
\begin{eqnarray}
( \cos  \omega  {d \over d\omega } \;-\; {j \over \sin  \omega }
)\; f = \nonumber
\\
- {j + i \epsilon  \over c} \; \sin ^{j + 1} \omega \; (\cos
\omega )^{-i\epsilon } \; [\; c F ( a - 1/2 , b + 1/2 , c ; z ) \;
+ \nonumber
\\
+ (a - 1/2 - c) \; F(a - 1/2, b + 1/2, c + 1 ; z)\; ]  = \nonumber
\\
= - {j + i \epsilon  \over 2}\; \sin ^{j + 1} \omega \; (\cos
\omega )^{-i\epsilon }\; {j - i \epsilon  \over 2}\; F(a + 1/2, b
+ 1/2, c + 1; z) \; ]\;  ; \nonumber
\end{eqnarray}

\noindent and further get
\begin{eqnarray}
( \cos  \omega  {d \over d\omega } \;-\; {j \over \sin  \omega } )
\; f = = - {j^{2}  + \epsilon ^{2} \over 2 j  + 3 } \;
U_{-i\epsilon ,j +1} \; . \label{2.11}
\end{eqnarray}

Finally, for the case (\ref{2.8d}) we have
\begin{eqnarray}
( \cos  \omega  {d \over d\omega } \;+\; {j +1 \over \sin  \omega
} ) \; f = \nonumber
\\
= \sin ^{j-1} \omega \; (\cos \omega )^{-i\epsilon }\; [\; (2 j +
1)\; F \; + \; z \; (i \epsilon  - j)\; F \; + \; 2 z (1 - z) \;
{d \over dz} \; F \;  ] \; ; \nonumber
\end{eqnarray}

\noindent from whence, using the formula
\cite{1953-Bateman-Erdelyi}
\begin{eqnarray}
z {d \over dz}\; F(\alpha , \beta , \gamma  ; z)  = (\gamma  - 1)
\; [\; F(\alpha , \beta  , \gamma + 1 ; z ) \; -\; F(\alpha ,
\beta , \gamma  ; z) \;] \nonumber
\end{eqnarray}

\noindent we get
\begin{eqnarray}
( \cos  \omega  {d \over d\omega } \;+\; {j +1 \over \sin  \omega
}  ) \;f = \sin ^{j-1} \omega \; (\cos \omega )^{-i\epsilon }
\times \nonumber
\\
\times [\; 2 (c - 1)\; (1 - z)\; F(a - 1/2, b + 1/2, c  + 1; z )\;
+ \nonumber
\\
+ \; 2 (c - a - 1/2) z\;  F (a - 1/2, b + 1/2, c ; z ) \; ] \; .
\nonumber
\end{eqnarray}

\noindent The term in square brackets equals to
$$
 (2 j + 1)  F(a - 1/2, b - 1/2, c - 1 ; z )\;,
 $$
  therefore  we have obtained
\begin{eqnarray}
 ( \cos  \omega  {d \over d\omega } \;+\;
{j +1 \over \sin  \omega }  ) \; f = ( 2 j  + 1 ) \; U_{-i\epsilon
,j-1} \;  . \label{2.12}
\end{eqnarray}

Collecting the results, we get

\begin{eqnarray}
\Pi- \mbox{wave} \; , \;\;\; ( A^{0} \;+\; A^{5})^{\Pi}  = e^{(-i
\epsilon +1 ) t } \; ( + j  - i \epsilon )\; U_{-i\epsilon +1,j}
\; Y_{jm} \; , \nonumber
\\
( A^{0} \;-\; A^{5})^{\Pi} = e^{(-i\epsilon - 1) t} \; ( - j - i
\epsilon ) \; U_{-i\epsilon -1,j}\; Y_{jm} \; , \nonumber
\\
{\bf  A}^{\Pi} = e^{-i\epsilon t} \;  [ \sqrt{{j + 1 \over 2j + 1
}} \; {\bf  Y}^{(j + 1)}_{jm}\; {- ( j^{2} + \epsilon ^{2}) \over
2 j + 3 } \; U_{-i\epsilon ,j + 1} \; - \nonumber
\\
- \; \sqrt{{j \over 2j + 1 }}\; {\bf  Y}^{(j-1)}_{jm}\; ( 2 j  + 1
) \; ] \; U_{-i\epsilon , j -1} \; ; \nonumber \label{2.13a}
\\
E- \mbox{wave}\; , \;\;\; {\bf  A}^{(E)} = e^{-i\epsilon t} \;  [
\; \sqrt{{j \over 2j + 1 }}\; {\bf  Y}^{(j + 1)}_{jm} \; {-( j^{2}
+ \epsilon ^{2}) \over 2 j + 3 } \; U_{-i\epsilon ,j+1} \; +
\nonumber
\\
+ \; \sqrt{{j +1 \over 2j + 1 }} \; {\bf  Y}^{(j-1)}_{jm}\; ( 2 j
+ 1 ) \; ] \; U_{-i\epsilon ,j-1} \; ; \nonumber \label{2.13b}
\\
M-\mbox{wave} \; , \;\;\; {\bf  A}^{(M)} = e^{-i\epsilon t} \;
{\bf  Y}^{(j)}_{jm}(\theta ,\phi ) \; U _{-i\epsilon ,j}(z)    \;
. \label{2.13c}
\end{eqnarray}

\noindent Components  $A^{0}$  and $A^{5}$  for $M-,E$-waves can
be found from comparing vector parts ${\bf A}$ for these waves
with given by
 (\ref{13.2.28a})  --  (\ref{13.2.28d})
\begin{eqnarray}
\nu  = j \; , \qquad {\bf  A} = e^{-i\epsilon t} \; U_{-i\epsilon
,j}\; {\bf  Y}^{(j)}_{jm}(\theta ,\phi ) \;  , \;  h_{0} = 1 \; ,
\;\; A^{0} = 0 \;\; , \; \; A^{5} =  0 \;     ; \nonumber
\\
\nu  = j +1 \; , \qquad f_{0} = \sqrt{{j + 1 \over 2j + 1}} \; ,
\;\; {\bf  A} = e^{-i\epsilon t} \; {\bf  Y}^{(j+1)}_{jm} \;
U_{-i\epsilon ,j+1} \;\;  , \nonumber
\\
( A^{0} \pm A^{5}) = e^{(-i\epsilon \pm  1 ) t }\; Y_{jm} \; {2j +
3 \over - i \epsilon  }\; \sqrt{{j + 1 \over 2j + 1}} \; ;
\nonumber
\\
\nu = j-1  \;  , \qquad g_{0} = \sqrt{{j \over 2j + 1}} \; , \;\;
{\bf  A} = e^{-i\epsilon t} \; {\bf  Y}^{(j-1)}_{jm} \;
U_{-i\epsilon ,j-1}\;  , \; \nonumber
\\
( A^{0} \; \pm \; A^{5}) = e^{(-i\epsilon \pm 1)t}\; Y_{jm} \; [\;
{(j \mp i \epsilon )\; (j \mp i \epsilon + 1) \over i\epsilon  ( 2
j + 1 ) } \; \sqrt{{j + 1 \over 2j + 1}} \; ] \; U_{-i\epsilon \pm
1,j} \; . \nonumber
\\
\label{2.14c}
\end{eqnarray}

Let us compare  these  $(\Pi,M,E)$- and $(j, j \pm 1)$-waves. We
see that $M$- and $j$-solutions coincide. For $\Pi$ wave we see
for vector parts
\begin{eqnarray}
{\bf  A}^{(\Pi)}=   - \sqrt{{j + 1 \over 2j + 1}} \; {j^{2} +
\epsilon ^{2}\over 2 j + 3} \; {\bf  A}^{(j+1)} \;-\; \sqrt{{j
\over 2j + 1}}\; (2j + 1)\; {\bf  A}^{(j-1)}\;   ; \nonumber
\\
\label{2.15a}
\end{eqnarray}

\noindent also the equalities hold
\begin{eqnarray}
(A^{0} \pm A^{5})^{(\Pi)} =   - \sqrt{{j + 1 \over 2j + 1}} \;
{j^{2} + \epsilon ^{2} \over 2 j  + 3} \; (A^{0} \pm
A^{5})^{(j+1)}   - \nonumber
\\
 -
  \sqrt{{j \over 2j + 1}} (2j  + 1)\;
(A^{0} \pm A^{5})^{(j -1)} \;  . \nonumber \label{2.15b}
\end{eqnarray}

\noindent For $E$-wave we have
\begin{eqnarray}
{\bf  A}^{(E)} =  [\; -\sqrt{{j \over 2j + 1}} \; {j^{2} +
\epsilon ^{2} \over 2 j  + 3}\; {\bf  A}^{(j + 1)} \;+\; \sqrt{{j
+1 \over 2j + 1}}\; (2j + 1)\; {\bf  A}^{(j -1)} \; ] \; ;
\nonumber
\\
\label{2.16a}
\end{eqnarray}

\noindent from whence it follows $A^{0}$ and $A^{5}$  for
$E$-wave:
\begin{eqnarray}
(A^{0} \pm A^{5})^{(E)} = {j  - i \epsilon  \over i\epsilon } \;
\sqrt{j(j + 1)}\; e^{(-i\epsilon \pm 1)t}\;  Y_{jm}\;
 U_{-i\epsilon \pm 1,j} \; .
 \label{2.16b}
\end{eqnarray}

\section{Supplement: Correspondence principle in  de Sitter model}

Let us show that  solutions constructed in de Sitter  space are in
accordance with correspondence principle:
 namely, they give the know results in the limit of flat Minkowski  space.
Let specify the case of $\Pi$-wave.

Starting with  limiting identities
\begin{eqnarray}
\lim_{\rho  \rightarrow  \infty } [\; r^{j+1}\; U_{-i\epsilon
,j+1}\; ] = {2 \over \epsilon }^{p} \; \Gamma (1 + p ) \;
{J_{p}(\epsilon r) \over \sqrt{r}} \; ,\qquad p = j + 3/2  \; ,
\nonumber
\\
\lim_{\rho  \rightarrow  \infty  } [\;r^{j-1} \; U_{-i\epsilon
,j-1} \; ] = {2 \over \epsilon }^{p'}\; \Gamma (1 + p') \;
{J_{p'}(\epsilon r) \over \sqrt{r}} \; , \qquad p' = j - 1/2 \; ,
\nonumber
\end{eqnarray}

\noindent ($\rho$ is a curvature radius, $J_{\nu }(x)$ stands for
the Bessel function \cite{1953-Bateman-Erdelyi},
 we readily produce
\begin{eqnarray}
\lim_{\rho  \rightarrow  \infty  } [  r^{j-1}  {\bf  A}^{(\Pi)}  ]
=
 e^{-i\epsilon t}
\lim_{\rho  \rightarrow  \infty}
  [ \; - \sqrt{{j \over 2j + 1 }}  {j^{2} + \epsilon ^{2} \over \rho ^{2}
( 2 j + 3) }   \rho ^{j + 1} {\bf  Y}^{(j+1)}_{jm}   U_{-i\epsilon
,j+1}  - \nonumber
\\
- \sqrt{{j \over 2j + 1 }}   \rho ^{j-1}   {\bf  Y}^{(j-1)}_{jm}
U_{-i\epsilon ,j -1}   ]  = \nonumber
\\
= e^{ -i\epsilon t}\; [\; (2j + 1) \; \Gamma (j + 1/2) \; {2 \over
\epsilon }^{j-1/2} \;   ] \; {1 \over \sqrt{r}}\;\times \nonumber
\\
\times  [\; \sqrt{{j +1 \over 2j + 1 }}\; {\bf  Y}^{(j+1)}_{jm}\;
J_{j+3/2} (\epsilon r)  + \sqrt{{j \over 2j + 1 }} \; {\bf  Y}^{(j
-1)}_{jm} \; J_{j-1/2}(\epsilon r)
 ] ;
\nonumber
\end{eqnarray}

\noindent or with notation $ f_{\nu }(x) = \sqrt{\pi /2x} \;
J_{\nu +1/2}(x) $  we get
\begin{eqnarray}
\lim_{ \rho  \rightarrow  \infty  }  [   \rho ^{j-1} {\bf
A}^{(\Pi)}  ] \sim e^{-i\epsilon t}  \left  [  \sqrt{{j +1 \over
2j + 1 }} {\bf  Y}^{(j+1)}_{jm} \; f_{j+1}(\epsilon r) + \sqrt{{j
\over 2j + 1 }}  {\bf  Y}^{(j-1)}_{jm}  f_{j-1}(\epsilon r)
\right  ]  , \nonumber
\end{eqnarray}

\noindent which agrees with the known representation for $\Pi$
wave in flat space. Besides, bearing in ming identities
\begin{eqnarray}
\lim_{ \rho  \rightarrow  \infty  } [\;\rho ^{j} \; U_{-i\epsilon
\pm 1,j}\;] = {2 \over \epsilon }^{p} \; \Gamma (1 + p) \; {
J_{p}(\epsilon r) \over \sqrt{r}}\; ,\qquad p  = j  + 3/2   \; ,
\nonumber
\end{eqnarray}

\noindent we find  limiting form for $A^{0(\Pi)}$  and
$A^{5(\Pi)}$:
\begin{eqnarray}
\lim_{ \rho  \rightarrow  \infty  } [\;\rho ^{j-1}(A^{0} \pm
A^{5})^{(\Pi)}\;]  = \nonumber
\\
= \lim_{ \rho  \rightarrow  \infty  } \; \left [ \;  {\pm j - i
\epsilon  \rho  \over \rho }\; \exp  [(-i{E\rho  \over \hbar c}
\pm 1 ) {ct \over \rho } ] \; Y_{jm}(\theta ,\phi ) \; \rho ^{j}
\; U_{-i\epsilon \pm 1,j}\; \right   ]\;  ; \nonumber
\end{eqnarray}

\noindent from this it follows
\begin{eqnarray}
\lim_{ \rho  \rightarrow  \infty  } \; [\;\rho ^{j-1} A^{5(\Pi)}\;
] = 0 \; , \; \lim_{ \rho  \rightarrow  \infty  } \; [\; \rho
^{j-1} A^{0(\Pi)}\; ] \sim e^{-i\epsilon t} \; Y_{jm}(\theta ,\phi
) \; f_{j}(\epsilon r) \; . \nonumber \label{2.17b}
\end{eqnarray}

\noindent which agrees with the known representation for $\Pi$
wave in flat space \cite{1948-Stratton}.

\section{Acknowledgment}

Authors are grateful to participants of seminar of Laboratory of
theoretical physics of Institute of physics of National Academy of
Sciences of Belarus for discission and advices.

\end{document}